\documentclass[
amsmath,
amssymb,
amsfonts,
aps,
nofootinbib,
preprintnumbers,
prd,
reprint,
superscriptaddress
]{revtex4-2}

\usepackage{xcolor}
\definecolor{MyDarkBlue}{rgb}{0.15,0.25,0.45} 
\usepackage[
linktocpage=true,
hypertexnames=false,
colorlinks=true,
citecolor=MyDarkBlue,
linkcolor=MyDarkBlue,
urlcolor=MyDarkBlue,
pdfauthor={
    Leron Borsten,
    Branislav Jurco,
    Hyungrok Kim,
    Tommaso Macrelli,
    Christian Saemann,
    Martin Wolf
},
pdftitle={Tree-Level Color--Kinematics Duality from Pure Spinor Actions},
pdfsubject={hep-th},
breaklinks=true
]{hyperref}

\usepackage{amsthm}
\usepackage{mathtools}
\usepackage{bbm}
\usepackage{bm}
\usepackage{enumerate}
\usepackage{mathrsfs}
\usepackage{multirow}
\usepackage{tikz}
\usetikzlibrary{matrix,cd,arrows,decorations.markings,decorations.pathmorphing,patterns}
\tikzset{>=stealth}
\tikzcdset{arrow style=tikz}

\tikzset{
    aux/.style={dashed},
    dottedline/.style={dotted},
    gluon/.style={
        decorate, 
        draw=black,
        decoration={
            snake,
            post=lineto,
            post length=0pt,
            segment length=4,
            amplitude=0.9
        }
    },
    matter/.style={
    },
}
\usepackage{stmaryrd}
\usepackage{cleveref}
\usepackage{ifthen}
\usepackage{slashed}
\newcommand{\makecommand}[3]{%
    \foreach \i in #3 {%
        \expandafter\xdef\csname #1\i\endcsname{\noexpand#2{\unexpanded\expandafter{\i}}}%
    }%
}
\newcommand{\latinalphabet}{A,a,B,b,C,c,d,D,E,e,F,f,G,g,H,h,I,i,J,j,K,k,L,l,M,m,N,n,O,o,P,p,Q,q,R,r,S,s,T,t,U,u,V,v,W,w,X,x,Y,y,Z,z}
\makecommand{I}{\mathbbm}{\latinalphabet}
\makecommand{I}{\mathbbm}{{CP}}
\makecommand{bf}{\mathbf}{\latinalphabet}
\makecommand{bm}{\bm}{\latinalphabet}
\makecommand{ca}{\mathcal}{\latinalphabet}
\makecommand{fr}{\mathfrak}{\latinalphabet}
\makecommand{rm}{\mathrm}{\latinalphabet}
\makecommand{sf}{\mathsf}{\latinalphabet}
\makecommand{sf}{\mathsf}{{id,SU,SO,Spin,SL}}
\makecommand{fr}{\mathfrak}{{su,so,spin,sl}}
\makecommand{sc}{\mathscr}{\latinalphabet}
\makecommand{tt}{\mathtt}{\latinalphabet}
\newcommand\CN\caN 

\newcommand{\wave}{\mathop\square}

\newcommand{\tr}{\mathrm{tr}} 

\newcommand{\parder}[2][]{%
    \ifthenelse{\equal{#1}{}}{%
        \frac{\partial}{\partial #2}%
    }{%
        \frac{\partial #1}{\partial #2}%
    }%
}
\newcommand{\delder}[2][]{%
    \ifthenelse{\equal{#1}{}}{%
        \frac{\delta}{\delta #2}%
    }{%
        \frac{\delta #1}{\delta #2}%
    }%
}
\newcommand{\inner}[2]{\langle#1,#2\rangle}
\newcommand{\eand}{{~~~\mbox{and}~~~}}

\newtheoremstyle{breaknodot}
{\topsep}{\topsep}%
{\itshape}{}%
{\bfseries}{}%
{0pt}{\thmname{#1}\thmnumber{ #2.}~\thmnote{ \normalfont(#3)}}%
\theoremstyle{breaknodot}

\let\oldblacksquare\blacksquare
\newcommand{\BBox}{{\textcolor{gray}\oldblacksquare}} 

\newcommand{\BVbox}{\text{BV}^{\BBox}}

\newcommand{\BVboxx}{\text{BV}^{\wave}}

\begin{document}
    
    \preprint{DMUS--MP--23/05}
    \preprint{EMPG--23--04}
    \title{Tree-Level Color--Kinematics Duality from Pure Spinor Actions}
    \author{Leron Borsten}
    \email[]{l.borsten@herts.ac.uk}
    \affiliation{Department of Physics, Astronomy and Mathematics, University of Hertfordshire\\ Hatfield, Hertfordshire, AL10 9AB, United Kingdom}
    \author{Branislav Jur{\v c}o}
    \email[]{branislav.jurco@gmail.com}
    \affiliation{Charles University Prague\\ Faculty of Mathematics and Physics, Mathematical Institute\\ Prague 186 75, Czech Republic}
    \author{Hyungrok Kim}
    \email[]{hk55@hw.ac.uk}
    \affiliation{Maxwell Institute for Mathematical Sciences\\ Department of Mathematics, Heriot--Watt University\\ Edinburgh EH14 4AS, United Kingdom}
    \author{Tommaso Macrelli}
    \email[]{tmacrelli@phys.ethz.ch}
    \affiliation{Department of Physics, ETH Zurich\\ 8093 Zurich, Switzerland}
    \author{Christian Saemann}
    \email[]{c.saemann@hw.ac.uk}
    \affiliation{Maxwell Institute for Mathematical Sciences\\ Department of Mathematics, Heriot--Watt University\\ Edinburgh EH14 4AS, United Kingdom}
    \author{Martin Wolf}
    \email[]{m.wolf@surrey.ac.uk}
    \affiliation{School of Mathematics and Physics, University of Surrey\\ Guildford GU2 7XH, United Kingdom}
    
    \date{\today}
    
    \begin{abstract}
        We prove that the tree-level scattering amplitudes for (super) Yang--Mills theory in arbitrary dimensions and for M2-brane models exhibit color--kinematics (CK) duality. Our proof for Yang--Mills theory substantially simplifies existing ones in that it relies on the action alone and does not involve any computation; the proof for M2-brane models establishes this result for the first time. Explicitly, we combine the facts that Chern--Simons-type theories naturally come with a kinematic Lie algebra and that both Yang--Mills theory and M2-brane models are of Chern--Simons form when formulated in pure spinor space, extending previous work on Yang--Mills currents~\cite{Ben-Shahar:2021doh}. Our formulation also provides explicit kinematic Lie algebras for the theories under consideration in the form of diffeomorphisms on pure spinor space. The pure spinor formulation of CK-duality is based on ordinary, cubic vertices, but we explain how ordinary CK-duality relates to notions of quartic-vertex 3-Lie algebra CK-duality for M2-brane models previously discussed in the literature.
    \end{abstract}
    
    \maketitle
    
    \section{Introduction}\label{sec:intro}
    
    Color--kinematics (CK) duality~\cite{Bern:2008qj,Bern:2010ue,Bern:2010yg} is a remarkable hidden feature of certain perturbative quantum field theories that puts the kinematic spacetime structure on the same footing as internal gauge or flavor symmetries. This idea has manifold implications and applications as reviewed in~\cite{Carrasco:2015iwa,Borsten:2020bgv, Bern:2019prr,Adamo:2022dcm,Bern:2022wqg}.
    In particular, it is key to the famous double copy prescription~\cite{Bern:2008qj,Bern:2010ue,Bern:2010yg} that allows for the construction of gravitational scattering amplitudes from a particular parameterization of the scattering amplitudes of supersymmetric Yang--Mills (SYM) theory.
    
    CK-duality and the double copy were originally discovered using on-shell amplitude technology. Having been shown the way by scattering amplitudes, we are naturally led to ask if we can return to the standard Lagrangian field theory starting point, possibly shedding further light on CK-duality~\cite{Bern:2010yg, Tolotti:2013caa, Borsten:2020zgj,Borsten:2021hua,Borsten:2021gyl,Borsten:2022vtg,Ben-Shahar:2022ixa}. This is all the more natural from the homotopy algebraic perspective, which puts Lagrangians and on-shell amplitudes on equal footing~\cite{Jurco:2018sby,Macrelli:2019afx,Arvanitakis:2019ald,Jurco:2019yfd}. In certain cases the kinematic Lie algebra and CK-duality are indeed symmetries of the Lagrangian, just as gauge invariance  is manifest in the Yang--Mills action~\cite{Borsten:2022vtg}.
    
    The main ingredient in our discussion is the notion of $\BVbox$-algebras introduced in~\cite{Reiterer:2019dys} for first-order Yang--Mills theory, where $\BBox$ denotes a second order differential operator. Any field theory with an underlying $\BVbox$-algebra has a kinematic Lie algebra~\cite{Borsten:2022vtg}, see also~\cite{Reiterer:2019dys}. That is, the tree-level Feynman diagram expansion is built from Lie-algebra valued vertices tied together with propagators $\frac{1}{\BBox}$. In the special case that $\BBox=\wave$, the spacetime d'Alembertian, this implies conventional CK-duality if one can extract finite numerators.\footnote{See also~\cite{Bonezzi:2022yuh,Bonezzi:2022bse}  for related work  manifesting CK-duality and applying it to the explicit construction of double field theory.}
    
    Identifying a theory's $\BVbox$-algebra (if one exists) is non-trivial, but there is an archetypal theory with $\BVbox$-algebra: Chern--Simons (CS) theory, cf.~\cite{Borsten:2022vtg,Ben-Shahar:2021zww}. When looking for CK-dual field theories, we are thus led to consider theories with CS-like reformulations. There are two evident families of candidates\footnote{We do not have anything to say about cubic harmonic superspace actions~\cite{Galperin:2001uw,Schwab:2013hf}, which also take a Chern--Simons form, cf.~\cite{Buchbinder:2008vi}.}: holomorphic CS theory on twistor space (reviewed in~\cite{Wolf:2010av,Adamo:2013cra}) and pure spinor CS actions (reviewed in~\cite{Bedoya:2009np,Cederwall:2010wf,Hoogeveen:2010aa,Cederwall:2013vba,Berkovits:2017ldz,Eager:2021wpi,Cederwall:2022fwu}). For the former, one obtains $\BVbox$-algebras for self-dual and full SYM theory with $\BBox=\wave$ in the self-dual case, but a more complicated expression in the full case~\cite{Borsten:2022vtg}. 
    
    Here, we consider pure spinor actions.\footnote{The pure spinor formalism has been previously applied to the study of CK-duality of SYM theory in e.g.~\cite{Mafra:2011kj,Mafra:2014gja,Mafra:2015mja,Bridges:2019siz}. It would be interesting to develop the relation to the present discussion.} For SYM theory, the pure spinor action is already in CS form, and its propagator suitably induces a conventional form of CK-duality as first observed in~\cite{Ben-Shahar:2021doh}, which shows that Berends--Giele currents of SYM theory come with a kinematic Lie algebra. However, a problem arises when turning these currents into CK-dual kinematic numerators of scattering amplitudes:  the \emph{tree-level} scattering amplitudes require an integral over pure spinor space that generically diverges for individual diagrams. One may skirt the divergences by regularizing the $\sfb$-operator~\cite{Berkovits:2006vi}, but the regularized $\sfb$-operator fails to be of second order as required for CK-duality.
    
    We circumvent this issue using the $Y$-formalism~\cite{Matone:2002ft,Oda:2005sd,Oda:2007ak} to demonstrate tree-level CK-duality of SYM theory directly from the action. Our new proof simplifies the existing one and exposes much more clearly underlying algebraic structures, such as the $\BVboxx$-algebra and the kinematic Lie algebra. We then identify the notion of $\BVbox$-module that governs kinematic Lie algebras for gauge--matter theories, cf.~\cite{Johansson:2014zca} for the discussion of CK-duality in this context. As an important application of this formalism, we show that the pure spinor actions for the Bagger--Lambert--Gustavsson (BLG), Aharony--Bergman--Jafferis--Maldacena (ABJM), and Aharony--Bergman--Jafferis (ABJ) models of~\cite{Cederwall:2008vd,Cederwall:2008xu} (cf.~\cite{Cederwall:2009ay}) imply suitable $\BVboxx$-algebras to establish all-order tree-level CK-duality. This completes partial results on CK-duality of Chern--Simons--matter (CSM) theories in the literature~\cite{Bargheer:2012gv,Huang:2012wr,Huang:2013kca,Sivaramakrishnan:2014bpa,Ben-Shahar:2021zww}. Our CK-duality for these theories uses cubic vertices; we explain how this implies a 3-Lie algebraic CK-duality using quartic vertices~\cite{Bargheer:2012gv,Huang:2012wr,Huang:2013kca,Sivaramakrishnan:2014bpa}.
    
    We will be very concise in our definition of the mathematical tools we use. A much more detailed exposition of the mathematical background is found in~\cite{Borsten:2023ned}.
    
    \section{Algebras underlying CK-duality}\label{sec:CK-algebra}
    
    To concisely encode CK-duality from the perspective of an action, we need to reformulate the action in order to make an additional algebraic structure visible. 
    
    In particular, we rewrite the action into an equivalent form with only cubic interaction terms. This can be achieved by introducing auxiliary fields, blowing up non-cubic interaction vertices into cubic ones as done previously, e.g., in~\cite{Bern:2010yg,Tolotti:2013caa,Anastasiou:2018rdx,Borsten:2021hua}. The Batalin--Vilkovisky (BV) formalism then produces a vector space graded by ghost number and a BV-differential encoding gauge transformations and equations of motion. The resulting structure dualizes (cf.~e.g.~\cite{Jurco:2018sby,Borsten:2021hua}) to a differential graded (dg) Lie algebra whose differential encodes the linearized gauge transformations and the linearized action, and whose Lie bracket encodes the non-linear corrections to both.\footnote{More generally, the BV complex of an arbitrary action dualizes to an $L_\infty$-algebra, with potential terms that are monomials of degree $d$ giving rise to $(d-1)$-ary brackets~\cite{Jurco:2018sby,Jurco:2019bvp}.}
    
    This dg-Lie algebra factorizes into the (ungraded) gauge Lie algebra and a dg-commutative algebra~\cite{Zeitlin:2008cc,Borsten:2021hua}. Such a factorization is possible\footnote{with dg-Lie algebras and dg-commutative algebras replaced by $L_\infty$- and $C_\infty$-algebras, respectively} for any ordinary gauge theory and amounts to the familiar color-stripping in the physics literature. 
    
    CK-duality then appears as a refinement of the dg-commutative algebra structure. In our previous work~\cite{Borsten:2020zgj,Borsten:2021hua,Borsten:2021gyl}, this appeared as compatibility with a twisted tensor product. Here, we focus on the notion of $\BVbox$-algebra~\cite{Reiterer:2019dys,Borsten:2022vtg}, see also~\cite{Akman:1995tm,Borsten:2022ouu,Borsten:2023ned}. A $\BVbox$-algebra is a dg-commutative algebra, i.e.~a $\IZ$-graded vector space endowed with a differential $Q$ and a graded-commutative product $\sfm(-,-)$, together with an operator $\sfb$ of degree $-1$ such that $\sfb^2=0$ and $\sfb$ is a second-order differential operator (in the sense of~\cite{koszul1985crochet,Akman:1995tm}). We usually denote the anticommutator $Q\sfb+\sfb Q$ in a $\BVbox$-algebra by $\BBox$. 
    
    The fact that $\sfb$ is of second order implies that the \emph{derived bracket}
    \begin{equation}\label{eq:derived_bracket}
        \{\phi,\psi\}\coloneqq\sfb\sfm(\phi,\psi)-\sfm(\sfb\phi,\psi)-(-1)^{|\phi|}\sfm(\phi,\sfb\psi)
    \end{equation}
    for $\phi,\psi$ color-stripped fields of ghost numbers $|\phi|,|\psi|$ is a (grade-shifted) Lie bracket, see e.g.~\cite[Proposition 1.2]{Getzler:1994yd}. This bracket turns out to be precisely the kinematic Lie algebra that combines with the gauge Lie algebra in cubic vertices. A kinematic Lie algebra then implies conventional CK-duality if $\BBox=\wave$, the spacetime d'Alembertian, and if the resulting numerators are finite up to distributional factors implementing momentum conservation.
    
    As an immediate example, consider plain CS theory on a three-dimensional spacetime $M$ for connections in a topologically trivial principal bundle $P=M\times \sfG$ for $\sfG$ some Lie group with Lie algebra $\frg$. The dg-Lie algebra reads as $\frL=\frg\otimes \Omega^\bullet(M)$, where $\Omega^\bullet(M)$ denotes the dg-commutative algebra given by the de Rham complex consisting of the differential forms on $M$. That is, $Q$ is the exterior derivative and the commutative product is the wedge product, $\sfm(-,-)=\wedge$. While the scattering amplitudes of asymptotically free fields are trivial, we can consider amplitudes for harmonic one-forms, as done, e.g., in~\cite{Ben-Shahar:2021zww} or~\cite{Borsten:2022vtg}. A natural choice for the operator $\sfb$ is then $\sfb=-\rmd^\dagger$ the codifferential defined with respect to a metric on $M$. This is indeed a second order differential operator in the sense of~\cite{koszul1985crochet,Akman:1995tm}, and we obtain a $\BVboxx$-algebra as $Q\sfb+\sfb Q=\wave$. The resulting kinematic Lie algebra is the Schouten--Nijenhuis algebra of multivector fields~\cite{Borsten:2022vtg}, and the scattering amplitudes of harmonic one-forms exhibit CK-duality as first noted in~\cite{Ben-Shahar:2021zww}. 
    
    \section{Supersymmetric Yang--Mills theories}
    
    \subsection{Pure spinor formulation}
    
    We first explain how the pure spinor approach for SYM theory manifests CK-duality, building on the observations for currents of~\cite{Ben-Shahar:2021doh}.
    
    The non-minimal pure spinor space $M_{\text{10D}\,\caN=1}$ of ten-dimensional SYM theory~\cite{Berkovits:2005bt} enlarges the usual superspace $\IR^{10|16}$ to a supermanifold coordinatized by $(x^M,\theta^A,\lambda^A,\bar\lambda_A,\rmd\bar\lambda_A)$, where indices $A$ belong to the $\mathbf{16}$ or $\overline{\mathbf{16}}$ of $\sfSpin(1,9)$. Note that $\rmd\bar \lambda_A$ is to be considered as an independent variable from $\bar \lambda_A$, and we use this notation to follow the common conventions in the pure spinor literature. All coordinates are commuting except for the coordinates $(\theta^A,\rmd\bar\lambda_A)$, which are anticommuting. The coordinates $(\lambda^A,\bar\lambda_A,\rmd\bar\lambda_A)$ carry ghost numbers $(1,-1,0)$ respectively and obey the constraints
    \begin{equation}\label{eq:pure_constraint_10D}
        \lambda^A\gamma^M_{AB}\lambda^B=\bar\lambda_A\gamma^{M\,AB}\bar\lambda_B=\bar\lambda_A\gamma^{M\,AB}\rmd\bar\lambda_B=0~.
    \end{equation}
    The covariant superderivatives $D_A$ on $\IR^{10|16}$ satisfy
    \begin{equation}
        D_AD_B+D_BD_A=-2\gamma^M_{AB}\parder{x^M}~,
    \end{equation}
    and~\eqref{eq:pure_constraint_10D} implies that the operator 
    \begin{equation}\label{eq:defOfQ}
        Q=\lambda^AD_A+\rmd\bar\lambda_A\parder{\bar\lambda_A}
    \end{equation}
    squares to zero. The volume form $\Omega_{M_{\text{10D}\,\caN=1}}$ given in~\cite{Berkovits:2005bt} permits an action principle for a scalar superfield $\Psi$ on $M_{\text{10D}\,\caN=1}$ of ghost number $1$ that takes values in a gauge metric Lie algebra $(\frg,\inner{-}{-}_\frg)$,
    \begin{equation}
        S_{\text{10D}\,\caN=1}=\int\Omega_{M_{\text{10D}\,\caN=1}}\inner{\Psi}{Q\Psi+\tfrac13[\Psi,\Psi]}_\frg~.
    \end{equation}
    One can compute perturbative scattering amplitudes using the propagator $\frac{\sfb}{\wave}$ in the Siegel gauge $\sfb\Psi=0$, where the $\sfb$-operator carries ghost number $-1$ and satisfies
    \begin{equation}\label{eq:b-properties}
        Q\sfb+\sfb Q=\wave\eand \sfb^2=0~,
    \end{equation}
    cf.~\cite{Bjornsson:2010wm,Bjornsson:2010wu}. Table~\ref{tab:coordinatesOperators} summarizes the properties of all coordinates and operators.
    \begin{table}[ht]
        \begin{center}
            \begin{tabular}{c|cccc}
                & \multirow{2}{*}{$\sfSpin(1,9)$} & mass & Grassmann & ghost
                \\[-3pt]
                & & dimension & degree & number
                \\
                \hline
                $x$ & $\mathbf{10}$ & $-1\phantom+$ & $0$ & $\phantom{+}0\phantom+$
                \\
                $\theta$ & $\mathbf{16}$ & $-\frac12\phantom+$ & $1$ & $\phantom{+}0\phantom+$
                \\
                $\lambda$ & $\mathbf{16}$& $-\frac12\phantom+$ & $0$ & $\phantom{+}1\phantom+$
                \\
                $\bar\lambda$ & $\overline{\mathbf{16}}$ & $\phantom{+}\frac12\phantom+$ & $0$ & $-1\phantom+$
                \\
                $\rmd\bar\lambda$ & $\overline{\mathbf{16}}$ & $\phantom{+}\frac12\phantom+$ & $1$ & $\phantom{+}0\phantom+$
                \\[1pt]
                \hline
                $D$ & $\overline{\mathbf{16}}$ & $\phantom{+}\frac12\phantom+$ & $1$ & $\phantom{+}0\phantom+$
                \\
                $Q$ & $\mathbf{1}$ & $\phantom{+}0\phantom+$ & $1$ & $\phantom{+}1\phantom+$
                \\
                $\sfb$ & $\mathbf{1}$ & $\phantom{+}2\phantom+$ & $1$ & $-1\phantom+$
                \\
                \hline
                $\Psi$ & $\mathbf{1}$ & $\phantom{+}0\phantom+$ & $1$ & $\phantom{+}1\phantom+$
            \end{tabular}
            \caption{Properties of 10D coordinates and operators.}
            \label{tab:coordinatesOperators}
        \end{center}
    \end{table}
    
    The $\sfb$-operator is not unique (cf.~\cite{Berkovits:2013pla,Jusinskas:2013sha,Cederwall:2022qfn}), but there is a convenient Lorentz-invariant form 
    \begin{equation}\label{eq:bcovariant}
        \sfb_0=-\frac{\bar \lambda_A\gamma^{M\,AB}D_B}{2\lambda^C\bar\lambda_C}\parder{x^M}+\cdots~,
    \end{equation}
    which is commonly used, e.g.~in~\cite{Ben-Shahar:2021doh}. To argue tree-level CK-duality of amplitudes, however, we will see that we have to use the non-Lorentz-invariant $\sfb$-operator of the $Y$-formalism~\cite{Matone:2002ft,Oda:2005sd,Oda:2007ak}, 
    \begin{equation}\label{eq:def_b_10}
        \sfb_Y=-\frac{v_A\gamma^{M\,AB}D_B}{2\lambda^Cv_C}\parder{x^M}
    \end{equation}
    for some reference pure spinor $v$ with $v_A \gamma^{M\,AB}v_B=0$. This operator satisfies~\eqref{eq:b-properties} and imposes a kind of axial gauge along $v$.
    
    \subsection{Tree-level CK-duality}
    
    Let us consider the color-stripped\footnote{That is, simply a function on pure spinor superspace, as opposed to a Lie algebra-valued function. Correspondingly all operators are also stripped of their action on the Lie algebra, which was simply the identity for all relevant examples.} scalar superfield, which we will denote by the same symbol $\Psi$, and first review how CK-duality emerges at the formal level following~\cite{Ben-Shahar:2021doh}.

    Both choices~\eqref{eq:bcovariant} and~\eqref{eq:def_b_10} for $\sfb$ are second-order differential operators with respect to the pointwise product on pure spinor superspace and satisfy~\eqref{eq:b-properties}. Therefore, both enhance the dg-commutative algebra of fields on pure spinor space with $Q$ as differential and the pointwise product to a $\BVboxx$-algebra. In particular, the derived bracket 
    \begin{equation}
        \{\Phi,\Psi\}\coloneqq\sfb(\Phi\Psi)-(\sfb\Phi)\Psi-(-1)^{|\Phi|}\Phi\sfb\Psi
    \end{equation} 
    is a (shifted) Lie bracket, as explained in Section~\ref{sec:CK-algebra}, which, together with~\eqref{eq:b-properties} implies CK-duality of currents.
    
    Explicitly, the derived bracket 
    simplifies in Siegel gauge ($\sfb\Psi=0$) for external states, and we have 
    \begin{equation}
        \{\Phi,\Psi\}=\sfb(\Phi\Psi)~.
    \end{equation} 
    Since the propagator is $\sfb/\wave$, all internal lines on any Feynman diagram are also in Siegel gauge, and we can push $\sfb$ on to vertices so that each color-stripped vertex is given by the derived bracket 
    \begin{equation}
        \sfb(\Phi\Psi)=\{\Phi,\Psi\}~.
	\end{equation}
    Here, we see concretely that the color-stripped vertex is governed by Lie algebra structure constants, leading to a kinematic Lie algebra. Because the remaining propagator is $1/\wave$, we formally recover ordinary CK-duality for the currents. We note that the above argument has first been made for the covariant $\sfb$-operator~\eqref{eq:bcovariant} in~\cite{Ben-Shahar:2021doh}.
    
    Converting these currents to individual CK-dual tree-level numerators of amplitudes, however, involves an integral over $(\lambda,\bar\lambda)$ that suffers from two kinds of divergences, infrared (IR, $(\lambda^A,\bar\lambda_A)\to\infty$) and ultraviolet (UV, $(\lambda^A,\bar\lambda_A)\to0$). We should stress that these divergences appear in tree-level amplitudes, but are purely an artifact of the pure spinor formalism; in the end all tree-level amplitudes are, of course, finite. 
    
    The IR divergences are due to the non-compactness of pure spinor space. These can be regulated, following~\cite{Marnelius:1990eq,Berkovits:2005bt,Cederwall:2022fwu}, by inserting a $Q$-exact regulator into the measure, $\Omega_{M_{\text{10D}\,\caN=1}}\mapsto\Omega_{M_{\text{10D}\,\caN=1}}\rme^{-\epsilon\{Q,\chi\}} $, where $\epsilon$ is a real positive constant and $\chi$ is a pure spinor field of ghost degree $-1$. For $\chi=-\bar\lambda_A\theta^A+\cdots$ we have  
    \begin{equation}
        \rme^{-\epsilon\{Q,\chi\}}=\rme^{-\epsilon(\lambda^A\bar\lambda_A+\cdots)}~,
    \end{equation}
    where the first factor manifestly suppresses the would-be IR divergences~\cite{Berkovits:2006vi}. This clearly preserves the kinematic Lie algebra since the bracket is merely scaled.          
    
    More problematic are the $(\lambda^A,\bar\lambda_A)\to0$ UV divergences. For the covariant non-minimal formalism, the amplitude integrands contain singularities of the form $1/(\lambda^A \bar\lambda_A)^n$ arising from the propagator $\sfb_0/\wave$ as well as the Siegel gauge condition, $\sfb_0\Psi=0$. Using the well-known Berkovits--Nekrasov regulator~\cite{Berkovits:2006vi}, these singularities can be regulated and  cancel in the total scattering amplitudes. 
    
    More specifically, when using the non-minimal formalism with a Lorentz-covariant $\sfb$-operator there exists a regulator that is $Q$-invariant and does not change the $Q$-cohomology classes~\cite{Berkovits:2006vi},
    \begin{equation}
        \sfb_\epsilon=\rme^{-\epsilon(w_A\bar w^A+\cdots)}\sfb_0~,
    \end{equation}
    where $w_A,\bar w^A$ are conjugate to $\lambda^A,\bar\lambda_A$. Since $w_A,\bar w^A$ are conjugate momenta, this superficially spoils the second-orderness of the $\sfb$-operator. However, all we will require in the end is that the difference between this $\sfb$-operator and the one we will use later is $Q$-exact, which is the case. What is crucial in the context of CK-duality is that at tree-level the UV singularities are integrals of $Q$-exact terms; these must vanish due to the gauge invariance of the total amplitudes. This is also made explicit in~\cite{Ben-Shahar:2021doh}, where it is inductively proven for SYM theory and illustrative examples at low points are given.   
    
    This conclusion applies equally to the $Y$-formalism $\sfb$-operator~\cite{Oda:2007ak}. First, the $Y$-formalism $\sfb_Y$ is equivalent to the covariant $\sfb$-operator on the $Q$-cohomology~\cite{Oda:2007ak}: $\sfb_0\Psi=\sfb_Y\Psi$ for all representatives $\Psi$ of the $Q$-cohomology. Hence, since all singular contributions to the total amplitude are $Q$-exact in the covariant $\sfb$-operator formalism, it follows that all singular contributions are $Q$-exact in the $Y$-formalism. In conclusion, finite tree-level scattering amplitudes can be computed using the $Y$-formalism and all potential singularities sum into a $Q$-exact term that vanish due to gauge-invariance. This latter observation is all that will be needed for CK-duality in the $Y$-formalism. 
    
    Note first, however, that the preceding argument does not prevent \emph{individual} Feynman diagrams, and hence their numerators, from having singular terms. In the Lorentz-invariant nonminimal formalism using $\sfb_0$~\cite{Berkovits:2005bt}, these singularities are potentially problematic for CK-duality~\cite{Ben-Shahar:2021doh}. We shall briefly review the potential obstruction to CK-duality below, but it is helpful to first consider the analogous situation in the $Y$-formalism.  
    
    First note that we obtain $Y$-formalism Siegel gauge physical states (unintegrated vertex operators) by starting with the nonsingular representatives $\lambda^A\lambda^B\caA_{AB}$ of the antifield cohomology classes and applying $\sfb_Y$ to them~\cite{Aisaka:2009yp},
    \begin{equation}
        \begin{split}
            \Psi&=\sfb_Y(\lambda^A\lambda^B\caA_{AB})
            \\
            &=-\frac{v_A\gamma^{M\,AB}D_B}{2\lambda^A v_A}\parder{x^M}(\lambda^C\lambda^D\caA_{CD})~.
        \end{split}
    \end{equation}
    The singularities of external states and Feynman diagrams are thus of the form $1/(\lambda^Av_A)^n$.
    
    The kinematic Jacobi identities hold order by order in $1/\lambda^Av_A$, but need to be regulated. Again, in the total scattering amplitude, divergent terms from each diagram either cancel or combine into $Q$-exact terms and thus become discardable. However, we would like to discard the singular terms in each individual diagram, before summing into a $Q$-exact term, so as to regulate the individual numerators in a minimal-subtraction-like scheme. The potential worry is that \emph{if} $Q$ can change the degree of divergence, then, in principle, finite terms from each diagram might be required to form the singular $Q$-exact term in the sum. In this case, when minimally subtracting the singular terms in each diagram individually, these finite terms would also need to be dropped. This will change the finite part of the numerators and so potentially break CK-duality. However, since the operator $Q$, being independent of $v$, does not affect the degree of singularity near $\lambda^Av_A=0$, the terms in the numerators that must be discarded may be restricted to singular terms only. 
    
    More explicitly, let us split each Feynman diagram, $\gamma_i$, into three terms,
    \begin{equation}
        \gamma_i=\gamma_i^{0}+\gamma_i^{Q,\text{finite}}+\gamma_i^{Q,\text{finite}}~,
    \end{equation}
    where the finite, $\gamma_i^{Q,\text{finite}}$, and singular, $\gamma_i^{Q,\text{singular}}$, terms contribute to the $Q$-exact part of the total amplitude integrand
    \begin{equation}
        I=I^0+Q\Lambda~, 
    \end{equation}
    where
    \begin{equation}
        Q\Lambda=\sum_i\left(\gamma_i^{Q,\text{finite}}+\gamma_i^{Q,\text{singular}}\right).
    \end{equation}
    Now, because $Q$ preserves the degree of singularity near $\lambda^A v_A=0$, $\sum_i\gamma_i^{Q, \text{finite}}$ and $\sum_i \gamma_i^{Q, \text{singular}}$ are separately $Q$-exact. Consequently, one can simply drop   $\gamma_i^{Q, \text{singular}}$ in each diagram separately, while preserving the total amplitude. Since CK-duality holds order by order in $1/(\lambda^A v_A)^n$, the resulting `minimally-subtracted' numerators obey the kinematic Jacobi identities. 
    
    In summary, we can truncate away the singular terms in the numerators without losing kinematic Jacobi identities, similar to minimal subtraction in dimensional regularization. The minimally subtracted numerators provide a CK-dual parameterization of the scattering amplitudes with finite numerators.
    
    Therefore, we have all-order tree-level CK-duality for 10D SYM theory. By dimensional reduction and embedding nonmaximally SYM tree diagrams into maximal ones~(cf.~\cite{Chiodaroli:2013upa}), this establishes tree-level CK-duality for all pure YM theories with arbitrary amounts of supersymmetry in any dimension.
    
    Let us now return to the Lorentz-covariant formalism. Here, the UV divergence occurs near $\lambda^A\bar\lambda_A=0$ for each numerator. As before, in the total amplitude all such singularities combine into a $Q$-exact term that vanishes due to gauge invariance. However, $Q$ \emph{does} change the degree of singularity near $\lambda^A\bar\lambda_A=0$ and so we cannot run the argument presented above for the $Y$-formalism. As a consequence, the cancellation of singularities in the total scattering amplitude may require  both singular and nonsingular terms from individual diagrams combining to form $Q$-exact terms that vanish under integration. As this discards some non-singular parts of the numerators, this may ruin kinematic Jacobi identities of minimally subtracted numerators. 
    
    \section{General gauge--matter theories}
    
    \subsection{3-algebraic formulation}
    
    A gauge theory with matter that features CK-duality comes with an extension of the algebraic structures discussed above, as we explain in the following. Because matter fields take values in a representation of the gauge Lie algebra, we require an appropriate notion of Lie modules.
    
    A \emph{metric Lie module} $(\frg,\inner{-}{-}_\frg,V,\inner{-}{-}_V)$ consists of a metric Lie algebra $(\frg,\inner{-}{-}_\frg)$ with
    a real orthogonal $\frg$-representation $(V,\inner{-}{-}_V)$. Any metric Lie module has a product $\wedge\colon V^2\rightarrow\frg$ defined by
    \begin{equation}
        \inner{X}{u\wedge v}_\frg=\inner{u}{X\cdot v}_V
    \end{equation}
    for all $u,v\in V$ and $X\in\frg$, that is antisymmetric and $\frg$-equivariant.
    \begin{proof}
        For all $u,v\in V$ and $X,Y\in\frg$, anti-symmetry follows from $\inner{X}{u\wedge v}_\frg=\inner{u}{X\cdot v}_V=-\inner{X\cdot u}{v}_V=-\inner{v}{X\cdot u}_V=-\inner{X}{v\wedge u}_\frg$ and equivariance is due to $\inner{Y}{[X,u\wedge v]}_\frg=-\inner{[X,Y]}{u\wedge v}_\frg=-\inner{u}{[X,Y]\cdot v}_V=-\inner{u}{X\cdot(Y\cdot v)}_V+\inner{u}{Y\cdot(X\cdot v)}_V=\inner{X\cdot u}{Y\cdot v}_V-\inner{Y\cdot u}{X\cdot v}_V=\inner{X\cdot u}{Y\cdot v}_V-\inner{X\cdot v}{Y\cdot u}_V=\inner{Y}{(X\cdot u)\wedge v}_\frg-\inner{Y}{(X\cdot v)\wedge u}_\frg=\inner{Y}{(X\cdot u)\wedge v}_\frg+\inner{Y}{u\wedge(X\cdot v)}_\frg$.
    \end{proof}
    
    The $\frg$-equivariance
    \begin{equation}
        \inner{X\cdot(u\wedge v)}{Y}_\frg=\inner{(X\cdot u)\wedge v}{Y}_\frg+\inner{u\wedge(X\cdot v)}{Y}_\frg
    \end{equation}
    is, diagrammatically,
    \newcommand\nodearray{
        \matrix (m) [
        matrix of nodes,
        ampersand replacement=\&,
        column sep=0.1cm,
        row sep=0.1cm
        ]{
            $X$ \& {} \& {} \& {} \& $u$
            \\
            {} \& {} \& {} \& {} \& {}
            \\
            {} \& {} \& {} \& {} \& {}
            \\
            {} \& {} \& {} \& {} \& {}
            \\
            $Y$ \& {}\& {} \& {} \& $v$
            \\
        };
    }
    \begin{equation}\label{eq:Jacobi-mixed}
        \begin{tikzpicture}[
            scale=1,
            every node/.style={scale=1},
            baseline={([yshift=-.5ex]current bounding box.center)}
            ]
            \nodearray
            \draw [gluon] (m-1-1) -- (m-3-2.center);
            \draw [gluon] (m-5-1) -- (m-3-2.center);
            \draw [gluon] (m-3-2.center) -- (m-3-4.center);
            \draw [matter] (m-1-5) -- (m-3-4.center);
            \draw [matter] (m-5-5) -- (m-3-4.center);
        \end{tikzpicture}
        =
        \begin{tikzpicture}[
            scale=1,
            every node/.style={scale=1},
            baseline={([yshift=-.5ex]current bounding box.center)}
            ]
            \nodearray
            \draw [gluon] (m-1-1) -- (m-2-3.center);
            \draw [gluon] (m-5-1) -- (m-4-3.center);
            \draw [matter] (m-2-3.center) -- (m-4-3.center);
            \draw [matter] (m-1-5) -- (m-2-3.center);
            \draw [matter] (m-5-5) -- (m-4-3.center);
        \end{tikzpicture}
        +
        \begin{tikzpicture}[
            scale=1,
            every node/.style={scale=1},
            baseline={([yshift=-.5ex]current bounding box.center)}
            ]
            \nodearray
            \draw [gluon] (m-1-1) -- (m-3-4.center);
            \draw [gluon] (m-5-1) -- (m-3-2.center);
            \draw [matter] (m-3-2.center) -- (m-3-4.center);
            \draw [matter] (m-1-5) -- (m-3-2.center);
            \draw [matter] (m-5-5) -- (m-3-4.center);
        \end{tikzpicture}.
    \end{equation}
    The metrics $\langle-,-\rangle_\frg$ and $\langle-,-\rangle_V$ define a 3-bracket $\llbracket-,-,-\rrbracket\colon V^3\rightarrow V$ by
    \begin{equation}\label{eq:3-bracket-antisymmetric-defn}
        \inner{s}{\llbracket u,v,w\rrbracket}_V=\inner{s\wedge u}{v\wedge w}_\frg
    \end{equation}
    for all $s,u,v,w\in V$. If $\llbracket-,-,-\rrbracket$ is totally antisymmetric, $(V,\llbracket-,-,-\rrbracket)$ is a 3-Lie algebra in the sense of~\cite{Filippov:1985aa}. We call the above data a \emph{3-Lie algebra structure}.
    
    A gauge--matter theory with exclusively cubic interaction vertices is still described by a dg-Lie algebra, but this dg-Lie algebra will now be the sum of two components: the pure gauge part is again the tensor product of the gauge Lie algebra with a dg-commutative algebra, but the second, matter part will be the tensor product of a metric Lie module together with a differential graded \emph{metric Com module} with the matter--gauge interactions provided by the module structure. 
    
    The definition of (ungraded) metric Com modules mimics the above construction for commutative rather than Lie algebras. A metric Com module consists of a (possibly non-unital) metric commutative associative algebra $(\frC,\inner{-}{-}_\frC)$ (i.e.~$\inner{X}{Y}_\frC=\inner{Y}{X}_\frC$ and $\inner{XY}{Z}_\frC=\inner{X}{YZ}_\frC$ for all $X,Y,Z\in\frC$) with a symplectic $\frC$-module $(V,\inner{-}{-}_V)$, i.e.~a $\frC$-module $V$ with an $\frC$-invariant symplectic metric $\inner{-}{-}_V$. Then the product $\bullet\colon V^2\rightarrow\frC$ defined by
    \begin{equation}\label{eq:bulletProduct}
        \inner{X}{u\bullet v}_\frC=\inner{u}{X\cdot v}_V
    \end{equation}
    for all $u,v\in V$ and $X\in\frC$ is commutative and $\frC$-bilinear.
    \begin{proof}
        For all $X,Y\in\frC$ and $u,v\in V$, commutativity follows from $\inner{X}{u\bullet v}_\frC=\inner{u}{X\cdot v}_V=-\inner{X\cdot u}{v}_V=\inner{v}{X\cdot u}_V=\inner{X}{v\bullet u}_\frC$ and bilinearity is due to $\inner{Y}{X\cdot(u\bullet v)}_\frC=-\inner{XY}{u\bullet v}_\frC=-\inner{u}{X\cdot(Y\cdot v))}_V=\inner{X\cdot u}{Y\cdot v}_V=\inner{Y}{(X\cdot u)\bullet v}_\frC$.
    \end{proof}
    \noindent Analogously to the 3-bracket $\llbracket-,-,-\rrbracket$, we define here a 3-bracket $\llparenthesis-,-,-\rrparenthesis\colon V^3\to V$ by
    \begin{equation}\label{eq:3-bracket-symmetric-defn}
        \inner{s}{\llparenthesis u,v,w\rrparenthesis}_V=\inner{s\bullet u}{v\bullet w}_\frC
    \end{equation}
    for all $s,u,v,w\in V$.
    
    The preceding constructions generalize to the differential graded setting by inserting appropriate sign factors and requiring the evident compatibility of the differential with the products. 
    
    In the particular case of a theory with a 3-Lie algebra gauge structure and cubic interaction vertices, we find an underlying dg-metric Lie module that factors into a gauge metric Lie module and a dg-metric Com module, generalizing color-stripping to color--flavor-stripping. 
    
    \subsection{CK-duality}
    
    To capture CK-duality for a gauge--matter theory, we extend our $\BVbox$-algebra $\frB$ by an additional $\BVbox$-module describing the matter fields. Such a $\BVbox$-module is a dg-module $(V,Q_V)$ over $\frB$ (in the sense of dg-commutative algebras) which is endowed with a degree $-1$ map $\sfb_V\colon V\rightarrow V$ that squares to zero, is a second-order differential operator with respect to the module action of $\frB$ on $V$, and such that $Q_V\sfb_V+\sfb_VQ_V=\BBox$. 
    
    Just as a $\BVbox$-algebra $\frB$ comes with an associated kinematic Lie algebra in the form of the derived bracket~\eqref{eq:derived_bracket}, a $\BVbox$-module $(V,Q_V,\sfb_V)$ comes with an associated kinematic Lie module. By a trivial extension of the arguments in~\cite{Borsten:2022vtg}, cf.~also~\cite{Borsten:2023ned}, if a color-stripped theory with matter admits the structure of a $\BVbox$-module, it automatically enjoys gauge--matter CK-duality as long as $\BBox=\wave$ and the resulting numerators do not diverge.
    
    \section{M-brane models}
    
    \subsection{CK-duality of the BLG model}
    
    Following~\cite{Cederwall:2008xu}, consider the dimensional reduction of pure spinor superspace $M_{\text{10D}\,\caN=1}$ to three dimensions, obtaining the pure spinor superspace $M_{\text{3D}\,\caN=8}$ coordinatized by $(x^\mu,\theta^{\alpha i},\lambda^{\alpha i},\bar\lambda_{\alpha i},\rmd\bar\lambda_{\alpha i})$, where $\mu,\nu,\ldots=0,1,2$. Then the $\mathbf{16}$ of $\sfSpin(1,9)$ becomes the $\mathbf{2}\otimes\mathbf{8}$ of $\sfSpin(1,2)\times\sfSpin(7)$, so the spinor index $A$ splits into $(\alpha,i)$ with $\alpha,\beta=1,2$ (raised and lowered using $\varepsilon_{\alpha\beta}$) and $i,j=1,\ldots,8$ (raised and lowered using $\delta_{ij}$). The R-symmetry enlarges from $\sfSpin(7)$ to $\sfSpin(8)$. Indices $m,n=1,\ldots,8$ denote the vector representation $\mathbf{8_v}$ of $\sfSpin(8)$.
    
    Following~\cite{Cederwall:2008xu} further, we relax the ten-dimensional pure spinor constraints~\eqref{eq:pure_constraint_10D} in three dimensions to 
    \begin{equation}\label{eq:pure_constraint_3D}
        \lambda^{\alpha i}\gamma_{\alpha\beta}^\mu\lambda_i^\beta=\bar\lambda^{\alpha i}\gamma_{\alpha\beta}^\mu\bar\lambda^\beta_i=\bar\lambda^{\alpha i}\gamma_{\alpha\beta}^\mu\rmd\bar\lambda^\beta_i=0~.
    \end{equation}
    
    The color--flavor structure of the BLG model is a metric Lie module $(\frg,\langle-,-\rangle_\frg,V,\langle-,-\rangle_V)$ with $\frg=\frsu(2)\oplus\frsu(2)$ and $V=(\mathbf2,\mathbf1)\otimes(\mathbf1,\mathbf2)$. The metric $\langle-,-\rangle_\frg$ on $\frg$ has signature $(3,3)$, while $\langle-,-\rangle_V$ is positive-definite. The resulting 3-bracket~\eqref{eq:3-bracket-antisymmetric-defn} is totally antisymmetric.\footnote{Up to direct sums, this is the only finite-dimensional metric Lie module allowing for $\caN=8$ supersymmetry~\cite{Papadopoulos:2008sk,Gauntlett:2008uf,Hosomichi:2008jb,Schnabl:2008wj}.}
    
    The gauge multiplet of the BLG model belongs to a $\frg$-valued superfield $\Psi$ on $M_{\text{3D}\,\caN=8}$ of mass dimension~$0$ and ghost number $1$. The matter superfield $\Phi$ takes values in the tensor product of $V$ with the $\mathbf{8_v}$ of $\sfSpin(8)$ with mass dimension $\tfrac12$ and ghost number $1$. We quotient the matter field space by the relation
    \begin{equation}
        \Phi^m\sim\Phi^m+\lambda^{\alpha i}\gamma^m_{\alpha\beta}\rho^\beta_i
    \end{equation}
    for arbitrary $\rho^\alpha_i$. 
    
    The pure spinor action for the BLG model is then~\cite{Cederwall:2008xu}
    \begin{equation}\label{eq:BLGAction}
        \begin{aligned}
            S_{\text{3D}\,\caN=8}&=\int\Omega_{\text{3D}\,\caN=8}\Big(\inner{\Psi}{Q\Psi+\tfrac13[\Psi,\Psi]}_\frg
            \\
            &\kern1.5cm+g_{mn}\inner{\Phi^m}{Q_V\Phi^n+\Psi\Phi^n}_V\Big)
        \end{aligned}
    \end{equation}
    with $\Omega_{\text{3D}\,\caN=8}$ the appropriate volume form on $M_{\text{3D}\,\caN=8}$ as defined in~\cite{Cederwall:2008xu}, $g_{mn}\coloneqq\lambda^{\alpha i}\gamma_{mn}{}_\alpha{}^\beta\lambda_{i\beta}$, and $Q_V=Q$.
    
    Color--flavor-stripping the dg-metric Lie module underlying the action~\eqref{eq:BLGAction} yields a dg-metric Com module $V$ over $\frC$. For any $s,u,v,w\in V$, the product~\eqref{eq:bulletProduct} satisfies
    \begin{equation}
        \inner{s\bullet u}{v\bullet w}_\frC=\inner{s\bullet v}{u\bullet w}_\frC~,
    \end{equation}
    inducing a totally symmetric 3-bracket~\eqref{eq:3-bracket-symmetric-defn}. 
    
    There is a suitable $Y$-formalism $\sfb$-operator also here: picking a reference pure spinor $v$ with $v_{\alpha i} \gamma^{\mu\,\alpha\beta}\delta^{ij}v_{\beta j}=0$, define
    \begin{equation}\label{eq:def_b_3}
        \sfb=\sfb_V=-\frac{v_{\alpha i}\gamma^{\mu\,\alpha\beta}\delta^{ij}D_{\beta j}}{2\lambda^{\alpha i} v_{\alpha i}}\parder{x^\mu}~,
    \end{equation}
    which satisfies~\eqref{eq:b-properties} and is second-order with respect to the module action on the dg-metric Com module.  Table~\ref{tab:coordinatesOperators2} summarizes the properties of all objects.
    \begin{table}[ht]
        \begin{center}
            \begin{tabular}{c|cccc}
                & \multirow{2}{*}{$\sfSL(2,\IR)\times\sfSpin(8)$} & mass & Grassmann & ghost
                \\[-5pt]
                & & dimension & degree & number
                \\
                \hline
                $x$ & $\mathbf{(3,8_v)}$ & $-1\phantom+$ & $0$ & $\phantom{+}0\phantom+$
                \\
                $\theta$ & $\mathbf{(2,8_s)}$ & $-\frac12\phantom+$ & $1$ & $\phantom+0\phantom+$
                \\
                $\lambda$ & $\mathbf{(2,8_s)}$ & $-\frac12\phantom+$ & $0$ & $\phantom+1\phantom+$
                \\
                $\bar\lambda$ & $\mathbf{(2,8_c)}$ & $\phantom{+}\frac12\phantom+$ & $0$ & $-1\phantom+$
                \\
                $\rmd\bar\lambda$ & $\mathbf{(2,8_c)}$ & $\phantom{+}\frac12\phantom+$ & $1$ & $\phantom{+}0\phantom+$
                \\[1pt]
                \hline
                $D$ & $\mathbf{(2,8_s)}$ & $\phantom{+}\frac12\phantom+$ & $1$ & $\phantom{+}0\phantom+$
                \\
                $Q$ & $\mathbf{(1,1)}$ & $\phantom{+}0\phantom+$ & $1$ & $\phantom{+}1\phantom+$\\
                $\sfb$ & $\mathbf{(1,1)}$ & $\phantom{+}2\phantom+$ & $1$ & $-1\phantom+$
                \\
                \hline
                $\Psi$ & $\mathbf{(1,1)}$ & $\phantom{+}0\phantom+$ & $1$ & $\phantom+1\phantom+$
                \\
                $\Phi$ & $\mathbf{(1,8_v)}$ & $\phantom{+}\frac12\phantom+$ & $0$ & $\phantom{+}0\phantom+$
            \end{tabular}
            \caption{Properties of 3D coordinates and operators.}
            \label{tab:coordinatesOperators2}
        \end{center}
    \end{table}
    
    As in the case of SYM theory, the operator~\eqref{eq:def_b_3} induces a $\BVboxx$-algebra structure for the gauge part, which extends to a $\BVboxx$-module structure on the full dg-metric Com module. This establishes CK-duality for the currents of the BLG model based on \emph{cubic} vertices, not the quartic vertices anticipated by 3-Lie algebras.
    
    To turn currents into scattering amplitudes, one integrates expressions with singularities of the form $\frac{1}{\lambda^{\alpha i}v_{\alpha i}}$ over $(\lambda,\bar\lambda)$-space. It is then clear that our previous arguments regarding minimal subtraction of singularities still hold: we obtain all-order tree-level CK-duality for the BLG model.
    
    \subsection{ABJM/ABJ models with pure spinors}
    
    Some CSM theories with $\caN<8$ supersymmetry admit cubic pure spinor actions and thus enjoy tree-level CK-duality using the $Y$-formalism. These include the $\caN=6$ ABJM~\cite{Aharony:2008ug} and ABJ~\cite{Aharony:2008gk} models in the pure spinor formulation of~\cite{Cederwall:2008xu}. The pure spinor superspace $M_{\text{3D}\,\caN=6}$ for 3D $\caN=6$ theories results from truncating the $\sfSpin(8)$ R-symmetry to $\sfSpin(6)$, so $M_{\text{3D}\,\caN=6}\subset M_{\text{3D}\,\caN=8}$. The indices agree with those for the BLG model except that $k,l,m,n,p=1,\ldots,4$ denote the $\mathbf{4}$ of $\sfSpin(6)\cong\sfSU(4)$. After truncation, we have $\lambda^{\alpha mn}=-\lambda^{\alpha nm}$, but the properties of $Q$ and $\sfb$ are not affected; the volume form $\Omega_{\text{3D}\,\caN=6}$ remains dimensionless.
    
    The gauge algebra $\frg$ remains a metric Lie algebra, but the representation $V$ is a \emph{complex} $\frg$-representation since the matter fields are in the complex representation $\mathbf{4}$ of $\sfSpin(6)$. The pure spinor actions for ABJM and ABJ models are~\cite{Cederwall:2008xu} 
    \begin{equation}
        \begin{aligned}
            S_{\text{3D}\,\caN=6}&=\int \Omega_{\text{3D}\,\caN=6}\Big(\inner{\Psi}{Q\Psi+\tfrac13[\Psi,\Psi]}_\frg
            \\
            &\kern1.5cm+g^m{}_n\inner{\bar\Phi_m}{Q\Phi^n+\Psi\Phi^n}_V\Big)
        \end{aligned}
    \end{equation}
    with $g^m{}_n=\tfrac12\varepsilon_{\alpha\beta}\varepsilon_{klpn}\lambda^{\alpha mk}\lambda^{\beta lp}$.
    
    The kinematic vector space here does not admit a suitable symplectic metric without breaking the pure spinor formalism. We can, however, formally quadruple the matter field space such that the matter fields take values in $(V\oplus V^*)\otimes(\mathbf{4}\oplus \overline{\mathbf{4}})$. This violates the non-linear BRST symmetry~\cite{Cederwall:2008xu} and hence unitarity for arbitrary external states, but restricting to appropriate external states produces correct tree amplitudes.
    
    After this enlargement, the dg-Lie algebra factorizes into a (gauge) metric Lie module and a $\BVboxx$-module. Thus the ABJM and ABJ models are CK-dual at the tree level, in the usual sense, to all orders.

    \subsection{Relation to quartic CK-duality}\label{sec:relation-to-literature}
    
    Previous literature~\cite{Bargheer:2012gv,Huang:2012wr,Huang:2013kca,Sivaramakrishnan:2014bpa} (except~\cite{Ben-Shahar:2021zww}) considered CK-duality and double copy of CSM theories using 3-Lie algebras and quartic graphs instead of the usual cubic graphs. In this section, we explain the relation between the two notions.
    
    For the BLG model in the pure spinor formalism, the BRST symmetry requires total anti-symmetry and total symmetry of the 3-brackets $\llbracket-,-,-\rrbracket$ and $\llparenthesis-,-,-\rrparenthesis$, respectively~\cite{Cederwall:2008vd}. Integrating out auxiliary modes in $\Psi$ produces expected quartic and sextic vertices:
    \begin{equation}
        \Psi Q\Psi+\Phi\Psi\Phi+\Psi^3\mapsto \Phi^2 Q^{-1}\Phi^2+(Q^{-1}\Phi^2)^3+\cdots.
    \end{equation}
    One can use~\eqref{eq:Jacobi-mixed} and its dg-metric Com module analogue to rewrite sextic vertices using quartic ones:
    \renewcommand\nodearray{
        \matrix (m) [
        matrix of nodes,
        ampersand replacement=\&,
        column sep=0.1cm,
        row sep=0.1cm
        ]{
            {} \& {} \& {} \& {} \& {}
            \\
            {} \& {} \& {} \& {} \& {}
            \\
            {} \& {} \& {} \& {} \& {}
            \\
            {} \& {} \& {} \& {} \& {}
            \\
            {} \& {}\& {} \& {} \& {}
            \\
        };
    }
    \begin{equation}
        \begin{tikzpicture}[
            scale=1,
            every node/.style={scale=1},
            baseline={([yshift=-.5ex]current bounding box.center)}
            ]
            \nodearray
            \draw [matter] (m-1-2) -- (m-2-3.center) -- (m-1-4);
            \draw [gluon] (m-2-3.center) -- (m-3-3.center);
            \draw [gluon] (m-3-3.center) -- (m-4-2.center);
            \draw [gluon] (m-3-3.center) -- (m-4-4.center);
            \draw [matter] (m-5-2) -- (m-4-2.center) -- (m-4-1);
            \draw [matter] (m-5-4) -- (m-4-4.center) -- (m-4-5);
        \end{tikzpicture}
        \to
        \begin{tikzpicture}[
            scale=1,
            every node/.style={scale=1},
            baseline={([yshift=-.5ex]current bounding box.center)}
            ]
            \nodearray
            \draw [matter] (m-1-2) -- (m-3-2.center) -- (m-3-4.center) -- (m-1-4);
            \draw [gluon] (m-3-2.center) -- (m-4-2.center);
            \draw [gluon] (m-3-4.center) -- (m-4-4.center);
            \draw [matter] (m-5-2) -- (m-4-2.center) -- (m-4-1);
            \draw [matter] (m-5-4) -- (m-4-4.center) -- (m-4-5);
        \end{tikzpicture}
        +
        \begin{tikzpicture}[
            scale=1,
            every node/.style={scale=1},
            baseline={([yshift=-.5ex]current bounding box.center)}
            ]
            \nodearray
            \draw [matter] (m-1-2) -- (m-3-4.center) -- (m-3-2.center) -- (m-1-4);
            \draw [gluon] (m-3-2.center) -- (m-4-2.center);
            \draw [gluon] (m-3-4.center) -- (m-4-4.center);
            \draw [matter] (m-5-2) -- (m-4-2.center) -- (m-4-1);
            \draw [matter] (m-5-4) -- (m-4-4.center) -- (m-4-5);
        \end{tikzpicture}.
    \end{equation}
    The right two diagrams consist of quartic $\Phi^4$ vertices. By construction, the coefficients of the resulting total quartic vertex are totally antisymmetric. Equivariance of the cubic vertices induces equivariance and thus CK-duality of the quartic $\Phi^4$ vertices. For the BLG model, cubic CK-duality therefore implies quartic CK-duality, agreeing with the observation in~\cite{Huang:2012wr,Huang:2013kca} that on-shell 3-Lie algebra CK-duality of BLG holds at $\le10$ points with double copy to 3D $\caN=16$ supergravity.
    
    Next, consider the ABJM and ABJ models. Here, the 3-brackets~\eqref{eq:3-bracket-antisymmetric-defn} and~\eqref{eq:3-bracket-symmetric-defn} still exist but are not totally (anti-)symmetric (being merely cyclic with respect to the metric). Hence, when translating the cubic graphs into the corresponding quartic graphs, one must remember the cyclic order of the attached edges. Thus, the scattering amplitude is partitioned into terms labeled not by unadorned quartic trees but by quartic trees with extra labels. This accords with the observation in~\cite{Huang:2013kca,Sivaramakrishnan:2014bpa} that, for the ABJM model, the quartic BCJ identities and quartic double copy (with unadorned quartic graphs) fail.
    
    \section{Concluding remarks}
    
    Our observations imply that one can double-copy~\cite{Borsten:2020zgj,Borsten:2021hua} the pure spinor actions of YM theory to obtain pure spinor actions of 10D and 3D $\caN=16$ supergravity, using the formalism of~\cite{Borsten:2023ned}.
    
    We note that our claim of cubic CK-duality for $\caN=6$ CSM theories does not contradict the result of~\cite{Ben-Shahar:2021zww} that $\caN=4$ is the maximal supersymmetry for CSM theories compatible with CK-duality: in the latter paper, only adjoint matter is considered, while we allow for general matter.
    
    It is very important to stress that our arguments apply only to the tree level, and there are fundamental obstructions to reaching the loop level. Suppose a cubic action $S$ of (e.g.~maximally) SYM theory existed, potentially formulated on some auxiliary space (e.g.~pure spinors, twistor space, harmonic or projective superspace) that manifests CK-duality off shell for all fields in some gauge. Further assume that after Kaluza--Klein expanding in the auxiliary coordinates and integrating out all auxiliary fields, this action reproduces the standard SYM action $S_\text{std}=\int\tr(F^2)+\cdots$ in a local, polynomial, Lorentz-invariant gauge. Then, by assumption, the off-shell tree-level correlators of $S_\text{std}$, which equal those of $S$ with external legs restricted to $(c,A,\phi,\chi,A^+,\phi^+,\chi^+,c^+)$, are CK-dual. Further, $S_\text{std}$ computes SYM loop amplitudes correctly (with the standard path integral measure, i.e.\ defined using dimensional regularization etc.), and it can be truncated to the action of pure Yang--Mills theory $S_\text{std}^{\caN=0}$, which consists of all terms containing exclusively the gauge fields and the additional fields arising in the BV formalism. The action $S_\text{std}^{\caN=0}$ manifestly suffices for computing pure YM scattering amplitudes, both at tree and at loop level: these amplitudes can be glued out of off-shell pure YM tree correlators, which are a subset of the SYM off-shell tree correlators. Moreover, these amplitudes must be CK-dual because the tree correlators of our initial SYM theory are CK-dual by assumption. But this contradicts the result of~\cite{Bern:2015ooa} that arbitrary-dimensional pure Yang--Mills theory lacks loop-level CK-duality with Lorentz-invariant polynomial numerators compatible with Feynman rules. 
    
    From this perspective, the limitation $Q\sfb+\sfb Q=\BBox\neq\Box$ in the $\BVbox$-algebra structure identified using an ambitwistor action in~\cite{Movshev:2004ub,Mason:2005kn,Borsten:2022vtg} seems very natural. The $\BVbox$-algebra manifests a kinematic Lie algebra at both the tree and (with some mild assumptions) the loop level, without directly implying off-shell CK-duality.\footnote{This does not contradict~\cite{Bern:2015ooa}: the truncation to $\caN=0$ fails at the loop level due to a gauge anomaly.} 
    
    Similarly in the pure spinor picture, failure of loop CK-duality may be seen as an incompatibility between a regulator of an ultraviolet divergence (i.e.~the regulated $\sfb$-operator) and a tree-level symmetry (the kinematic algebra), or an anomaly, following the perspective of~\cite{Borsten:2021gyl,Borsten:2022vtg}.
    
    Altogether we conclude that, without further input, tree-level CK-duality is the best one can hope for to obtain from $\BVbox$-algebras underlying action principles for Yang--Mills theory.
    
    \
    
    \section*{Data Management}
    No additional research data beyond the data presented and cited in this work are needed to validate the research findings in this work. For the purpose of open access, the authors have applied a Creative Commons Attribution (CC BY) license to any Author Accepted Manuscript version arising.
    
    \
    
    \begin{acknowledgments}
        We thank Maor Ben-Shahar for discussions and detailed explanations of the results of~\cite{Ben-Shahar:2021doh} as well as Martin Cederwall for helpful comments. H.K. and C.S.~were supported by the Leverhulme Research Project Grant RPG-2018-329. B.J.~was supported by the GA\v{C}R Grant EXPRO 19-28628X.
    \end{acknowledgments}
    
    \bibliography{ref/bigone}
    
\end{document}